\documentclass{article}

\usepackage{arxiv}

\usepackage[utf8]{inputenc} 
\usepackage[T1]{fontenc}    
\usepackage{hyperref}       
\usepackage{url}            
\usepackage{booktabs}       
\usepackage{amsfonts}       
\usepackage{nicefrac}       
\usepackage{microtype}      
\usepackage{lipsum}		
\usepackage{algorithmic}
\usepackage{graphicx}
\usepackage{doi}
\usepackage{textcomp}
\usepackage{xcolor}

\usepackage{siunitx}
\usepackage{float}
\usepackage{tikz}
\usetikzlibrary{shapes.geometric,arrows}
\usepackage{pgfplots}
\usepackage{microtype}  
\usepgfplotslibrary{groupplots,dateplot}
\usetikzlibrary{patterns}
\pgfplotsset{compat=newest}

\newcommand\leqms[1]{$\leq \SI{#1}{\milli\second}$}

\title{Fine-Tuning MIDI-to-Audio Alignment using a Neural Network on Piano Roll and CQT Representations}

\author{ 
  Sebastian Murgul \\
	Klangio GmbH \\
	Karlsruhe, Germany \\
	\texttt{sebastian.murgul@klang.io} \\
	\And
  Moritz Reiser \\
	University of Music Karlsruhe \\
	Karlsruhe, Germany \\
	\texttt{moritz.reiser@t-online.de} \\
	\And
  Michael Heizmann \\
	Karlsruhe Institute of Technology \\
	Karlsruhe, Germany \\
	\texttt{michael.heizmann@kit.edu} \\
	\And
  Christoph Seibert \\
	University of Music Karlsruhe \\
	Karlsruhe, Germany \\
	\texttt{seibert@hfm-karlsruhe.de} \\
}

\renewcommand{\headeright}{}
\renewcommand{\undertitle}{}
\renewcommand{\shorttitle}{Fine-Tuning MIDI-to-Audio Alignment}

\hypersetup{
pdftitle={Fine-Tuning MIDI-to-Audio Alignment using a Neural Network on Piano Roll and CQT Representations},
pdfauthor={Sebastian Murgul, Moritz Reiser, Michael Heizmann, Christoph Seibert},
pdfkeywords={Music Information Retrieval, MIDI-to-Audio Alignment, Neural Network},
}

\begin{document}
\maketitle

\begin{abstract}
  In this paper, we present a neural network approach for synchronizing audio recordings of human piano performances with their corresponding loosely aligned MIDI files. The task is addressed using a Convolutional Recurrent Neural Network (CRNN) architecture, which effectively captures spectral and temporal features by processing an unaligned piano roll and a spectrogram as inputs to estimate the aligned piano roll. To train the network, we create a dataset of piano pieces with augmented MIDI files that simulate common human timing errors. The proposed model achieves up to 20\% higher alignment accuracy than the industry-standard Dynamic Time Warping (DTW) method across various tolerance windows. Furthermore, integrating DTW with the CRNN yields additional improvements, offering enhanced robustness and consistency. These findings demonstrate the potential of neural networks in advancing state-of-the-art MIDI-to-audio alignment.
\end{abstract}


\section{Introduction}\label{sec:introduction}


In the last decade, substantial progress has been made in the field of Music Information Retrieval (MIR) using deep learning techniques.
A key downstream task is Automatic Music Transcription (AMT) \cite{benetos_automatic_2018}, which seeks to convert audio recordings of music into a symbolic music representation like MIDI.
Since many of the approaches rely on frame-wise annotations and annotating real world recordings is quite time-consuming, the use of synthesized MIDI data has become popular. While this offers a viable path to obtaining large training datasets, incorporating real-world audio could lead to further improvements.
Using automatic MIDI-to-audio alignment can accelerate the annotation process and make real-world audio suitable for training.

Dynamic Time Warping (DTW) \cite{dannenberg_polyphonic_2003,hu_polyphonic_2003} is one of the most established algorithmic methods and is considered the industry standard. DTW has been proposed by Dannenberg and Hu in 2003 and uses dynamic programming to optimize a monotonic alignment path by minimizing the cost between the aligned feature vectors. These feature vectors consist of chroma features derived from the spectrogram of the audio and the piano roll of the MIDI. The original method has been improved by several extensions \cite{ewert_high_2009, pratzlich_memory_2016}, and it has proven to be extremely efficient for aligning or matching recordings of music to the corresponding MIDI transcriptions.
In 2009, Ewert et al.\ introduced the DLNCO features which improve the temporal accuracy without sacrificing the robustness \cite{ewert_high_2009}. The computational complexity of DTW is quadratic in the input length which becomes a major issue when dealing with longer audio recordings. Therefore, Prätzlich et al.\ proposed a Memory restricted Multiscale DTW (MrMsDTW) \cite{pratzlich_memory_2016}.
Besides the DTW method, Gingras et al.\ proposed a three-step matcher algorithm based on pitch clustering in 2011 \cite{gingras_improved_2011}.
In 2014, Chen et al.\ introduced two more algorithmic approaches \cite{chen_improved_2014}. The first one converts the note sequences into strings and performs a string matching. The second one uses the divide and conquer principle to split the input note sequences into smaller units that are synchronized recursively.
Besides the algorithmic approaches, neural networks have been used to improve the feature extraction and the DTW cost calculation.
Kwon et al.\ use an RNN-based music transcription algorithm in order to extract chroma features from an audio's spectrogram which can be used in a DTW alignment with chroma features extracted from MIDI \cite{kwon_audio_2017}.
Wang et al.\ use a triplet pair architecture to learn features from low-level representation using a hinge loss function \cite{wang_improvement_2019}.
Agrawal et al.\ propose a Siamese network architecture which performs a feature extraction on the audio spectrograms and then estimates a distance between the two audio inputs \cite{agrawal_learning_2021}.
A preprocessing pipeline incorporating DTW and a pre-trained transcription model is used in \cite{riley_high_2024} and \cite{maman_unaligned_2022} to align unaligned real-world transcriptions for fine-tuning.

In terms of AMT dataset creation, DTW is used for fine alignment of the MAESTRO dataset \cite{hawthorne_enabling_2018} in order to account for any jitter in the audio or MIDI recording. In Pop2Piano, Choi et al.\ use DTW to synchronize piano arrangements to the corresponding original music video \cite{choi_pop2piano_2023}.
One major downside of using these techniques for dataset alignment is that they only consider one complete timeframe at once and cannot perform a frequency band specific alignment. Therefore, it can occur that for example the lower notes of a chord are aligned perfectly, but the higher notes are annotated too early.

In this work, a deep learning approach for fine-tuning loosely aligned MIDI annotations, which could be derived from DTW, is introduced. 
The model is based on a CRNN architecture and is trained to estimate an aligned piano roll given an unaligned piano roll and the performance audio. 
For training, a suited methodology for creating synthetic unaligned training examples from the well known piano transcription dataset MAPS \cite{emiya_multipitch_2010} is presented.
Finally, the synchronization results are evaluated with real world examples as a proof of concept.
\section{Model Description}\label{sec:model}

The two critical components for our approach are the feature extraction of the two cross-modal inputs and the temporal modelling in order to determine the relation between the unaligned and aligned notes.

\subsection{Preprocessing}\label{sec:preprocess}

The MIDI files are converted to piano roll representations with pitches ranging from $21$ to $108$ and a temporal resolution of $100$ frames per second.
The velocity of the notes within the MIDI files is ignored in the piano roll creation, hence resulting in a binary image. 
In order to retain the individual note onsets of directly successive notes, small gaps with a width of one frame ($\SI{10}{\ms}$) are inserted before an onset if needed.
The input audio signal is resampled to $\SI{16}{\kHz}$. The CQT is calculated with a hop size of $160$ using the librosa library \cite{mcfee_librosa_2015}. To match the input piano roll, a total number of $88$ frequency bins with $12$ bins per octave is used, starting at a minimum frequency of $\SI{27.5}{\Hz}$. To highlight the contours in the CQT, the magnitude values are employed with a logarithmic scale.
Since the audio and the unaligned MIDI file can be of different durations, the CQT and the piano rolls are zero padded to a common temporal length.

\subsection{Network Architecture}

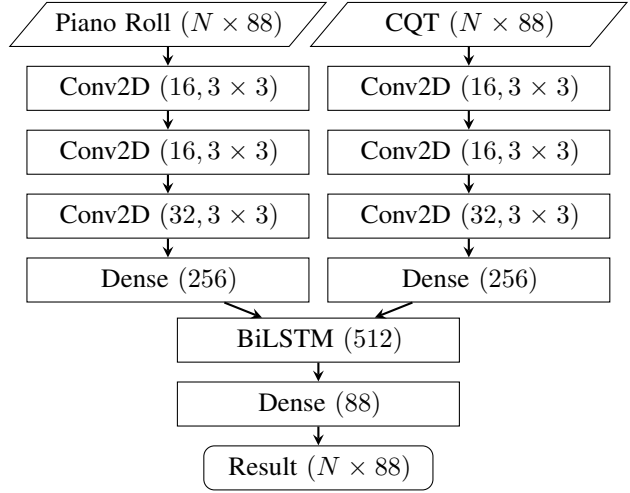
\begin{figure}[ht]
\tikzstyle{end} = [rectangle, rounded corners, text width=8em, minimum height=1em, text centered, draw]
\tikzstyle{line} = [thick,->,>=stealth]
\tikzstyle{input} = [trapezium, trapezium left angle=50, trapezium right angle=130, text width=8.5em, 	minimum height=1em, text centered, draw]
\tikzstyle{block} = [rectangle, draw, text width=10em, text centered, minimum height=1.6em]

\begin{tikzpicture}[node distance = 0.85cm, rotate=0, transform shape]
  \coordinate (dummy);
  \node [input, left of=dummy, node distance=2cm] (prStart) {Piano Roll $(N \times 88)$};
  \node [block, below of=prStart] (conv11) {Conv2D $(16, 3\times 3)$};
  \node [block, below of=conv11] (conv21) {Conv2D $(16, 3\times 3)$};
  \node [block, below of=conv21] (conv31) {Conv2D $(32, 3\times 3)$};
  \node [block, below of=conv31] (fc11) {Dense $(256)$};
  \node [input, right of=dummy, node distance=2cm] (cqtStart) {CQT $(N \times 88)$};
  \node [block, below of=cqtStart] (conv12) {Conv2D $(16, 3\times 3)$};
  \node [block, below of=conv12] (conv22) {Conv2D $(16, 3\times 3)$};
  \node [block, below of=conv22] (conv32) {Conv2D $(32, 3\times 3)$};
  \node [block, below of=conv32] (fc12) {Dense $(256)$};
  \node [block, below of=dummy, node distance=4.2cm] (lstm) {BiLSTM $(512)$};
  \node [block, below of=lstm] (fc2) {Dense $(88)$};
  \node [end, below of=fc2] (finish) {Result $(N \times 88)$};
  \draw [line] (prStart) -- (conv11);
  \draw [line] (conv11) -- (conv21);
  \draw [line] (conv21) -- (conv31);
  \draw [line] (conv31) -- (fc11);
  \draw [line] (fc11) -- (lstm);
  \draw [line] (cqtStart) -- (conv12);
  \draw [line] (conv12) -- (conv22);
  \draw [line] (conv22) -- (conv32);
  \draw [line] (conv32) -- (fc12);
  \draw [line] (fc12) -- (lstm);
  \draw [line] (lstm) -- (fc2);
  \draw [line] (fc2) -- (finish);
\end{tikzpicture}
  \centering
  \caption{Architecture of the proposed CRNN model for MIDI-to-audio alignment. Two parallel convolutional branches extract spectral and temporal features from the unaligned piano roll and CQT input. The features are concatenated and passed through a BiLSTM layer and dense output layer to estimate the aligned piano roll as a binary activation map over time and pitch.}
  \label{fig:architecture}
\end{figure}

The model architecture is visualized in Figure \ref{fig:architecture}.
It is organized in three blocks: two identical convolutional blocks for the feature extraction in the piano roll and the CQT images and one recurrent block that compares the extracted features in order to estimate an output piano roll.
The architecture of the convolutional blocks is inspired by the model for singing transcription by Schwabe et al.\ \cite{schwabe_dtmst_2022}. The CQT and the piano roll share the same dimensions of $N \times 88$ where $N$ represents the temporal dimension and $88$ corresponds to the number of pitches. These images are processed by three two-dimensional convolutional layers (Conv2D) with a kernel size of $3 \time 3$. The first two convolutional layers consist of 16 filters, the third one uses 32 filters. ReLU is used as activation function after the convolutional layers.
Max pooling of the pitch dimension with a size of $2$ is inserted after each of the last two convolutional layers.
The output of the last convolutional layer is fed into a fully connected layer (Dense). Batch Normalization is used after each convolutional layer to improve the feature stability and a dropout of 0.5 is employed after the intermediate Dense layers and the BiLSTM layer as a regularization method.
The results of both strands are concatenated to a shape of $N \times 512$ where $512$ corresponds to our embedding size. To improve the modelling of temporal constraints, the output is fed into a bidirectional LSTM layer (BiLSTM) with an input size of $512$ and a hidden size of $256$. The output of the BiLSTM is processed by another fully connected layer that returns a piano roll with the dimensions $N \times 88$. Using a sigmoid activation function allows the interpretation as probabilities.

\subsection{Postprocessing}\label{sec:postprocessing}

To construct the resulting MIDI sequence from the returned piano roll, a postprocessing step is employed.
Firstly, the piano roll is converted into a note sequence by thresholding the estimated piano roll with $0.5$ and segmenting the detected note blocks for each pitch by searching for transitions.
Since it is not guaranteed that the neural network returns the correct number of notes, the estimated notes are matched with the notes of the input MIDI file using the pitch and temporal order. After successful matching, the estimated notes' start and end times are used to update the input MIDI sequence. If a corresponding note is not found in the estimated sequence, the original unaligned note will be added to the resulting MIDI file instead. 
\section{Experiments}\label{sec:experiments}

This section outlines the experimental setup, including the datasets used, the training procedure, and the evaluation configuration employed to assess the proposed model.

\subsection{Datasets}\label{sec:dataset}

\subsubsection{Unaligned MAPS Dataset}\label{sec:maps}

To be able to train, validate, and test a neural network for synchronization, enough unsynchronized training data with dedicated labels is needed. Because such a dataset has not been available at the time of this work, the synthesized musical pieces of the MAPS database \cite{emiya_multipitch_2010} are used. For each of the pieces, the corresponding perfectly aligned MIDI file is augmented to obtain a loosely aligned version.
Firstly, the audio and MIDI files of each example are segmented into parts with a length of about $\SI{30}{\s}$ each, yielding more training data and a more consistent sequence length. The individual lengths can vary since points in time without active notes are searched in order to find reasonable splitting points.

The onsets and offsets of all notes are slightly shifted to simulate the timing variations characteristic of
human performance. Such timing variations can occur unintentionally, due to limitations in motor control or perceptual processes, but may also be deliberately employed as a means of musical expression or communication \cite{penel_timing_2004}.

\begin{figure}[htbp]
  \centering
  \providecommand{\fwidth}{0.48\textwidth}
\begin{tikzpicture}

  \definecolor{darkgray176}{RGB}{176,176,176}
  \definecolor{lightgray204}{RGB}{204,204,204}

  \begin{axis}[
      legend cell align={left},
      legend style={fill opacity=0.8, draw opacity=1, text opacity=1, draw=lightgray204, font=\small},
      legend pos = north west,
      tick align=outside,
      tick pos=left,
      width=\fwidth,
      height=0.6*\fwidth,
      axis on top,
      xlabel={Time in s},
      x grid style={darkgray176},
      xmin=-0.011609977324263, xmax=1.3815873015873,
      xtick style={color=black},
      xtick={0,0.3,0.6,0.9,1.2,1.5},
      xticklabels={$\num{0}$,$\num{0,3}$,$\num{0,6}$,$\num{0,9}$,$\num{1,2}$,$\num{1,5}$},
      y grid style={darkgray176},
      ylabel={Pitch},
      ymin=2.5, ymax=18.5,
      ytick style={color=black},
      ytick={0, 1, 2, 3, 4, 5, 6, 7, 8, 9, 10, 11, 12, 13, 14, 15, 16, 17, 18, 19, 20, 21, 22, 23, 24},
      yticklabels={ , , , , , $C4$, , , , , , , , , , , , $C5$}
    ]
    \addlegendimage{line legend,blue}\addlegendentry{Exact Note Length}
    \addlegendimage{line legend,red}\addlegendentry{Human Performance}
    \draw[draw=red,fill=red] (axis cs:0.34,4.5) rectangle (axis cs:1.11,5.5);
    \draw[draw=blue,fill=blue,very thin] (axis cs:0.3,4.9) rectangle (axis cs:1.05,5.1);
    \draw[draw=red,fill=red] (axis cs:0.38,8.5) rectangle (axis cs:1.01,9.5);
    \draw[draw=blue,fill=blue,very thin] (axis cs:0.3,8.9) rectangle (axis cs:1.05,9.1);
    \draw[draw=red,fill=red] (axis cs:0.24,11.5) rectangle (axis cs:1.13,12.5);
    \draw[draw=blue,fill=blue,very thin] (axis cs:0.3,11.9) rectangle (axis cs:1.05,12.1);

  \end{axis}

\end{tikzpicture}
  \caption{Illustration of timing deviations in a human piano performance. The blue bars represent ideal note lengths (as in a metronome-guided score), while the red bars show real-world timing deviations due to natural variation in note onsets and durations.}
  \label{fig:human_error_example}
\end{figure}
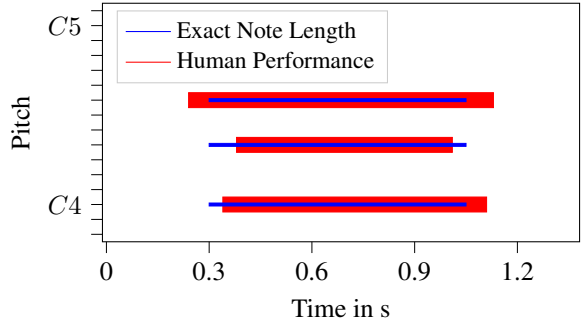

A possible reference for the maximum value of variation was derived from \cite{palmer_performance_1997} with $\SI{50}{\ms}$ for playing piano music and is used, for example, in \cite{devaney_estimating_2014} and \cite{simonetta_audiotoscore_2021}. However, given the complex interplay of perceptual, motor, and expressive factors influencing timing, and in order to accommodate a broader spectrum of timing deviations in our model, we chose to allow for a maximum onset/offset deviation of $\SI{100}{\ms}$. Figure \ref{fig:human_error_example} illustrates an example of such timing variations, showing how note boundaries are temporally displaced in human performance.

Besides shifting onsets and offsets, tempo fluctuations can also be simulated. Therefore, a scaling factor can be used to change the tempo of the unaligned MIDI files. This additional augmentation step is only applied for the experiments in Section \ref{sec:tempo}.

The resulting unaligned MAPS dataset consists of triplet examples, each containing an audio, a perfectly aligned MIDI, and an artificially unaligned MIDI file.
The dataset is split into three subsets. The $210$ synthesized examples are used as training split, and the $60$ real recordings of ENSTDkAm and ENSTDkCl are used as validation and test split respectively.

\subsubsection{Real World Piano Recordings}\label{sec:self_performed}

The unaligned MAPS dataset is created in a way that the audio performances are temporally accurate representations while the corresponding MIDI files contain additional human timing variations.
In the use case of recording a dataset however, the exact opposite is the case.
To test how the network deals with the task of adjusting MIDI scores to match the human performance, six real world piano performances have been recorded. These were played on a digital piano and recorded by a Digital Audio Workstation (DAW), allowing the saving of MIDI data for usage as perfectly aligned labels.
The temporally accurate MIDI file for each piece is taken from its digital musical score. During the recording of the pieces, a metronome was used, so only scenarios without tempo changes are considered here.


\subsection{Model Training}

To fit the model to the training data, the binary cross entropy loss is minimized using the Adam optimizer \cite{kingma_adam_2014} with a learning rate of $0.001$.
The training is performed using a mini-batch size of $8$ and is run for $50$ to $200$ epochs with early stopping.
The model converges quite fast resulting in a training time on an NVIDIA T4 with $\SI{16}{\giga\byte}$ RAM of $\SI{2}{\hour}$ on average.

\subsection{Evaluation Metrics}\label{sec:metrics}

For the evaluation of the model, the absolute onset errors and the alignment accuracy in different tolerance windows are calculated. The absolute onset errors are obtained by comparing the note onset times for corresponding notes in the estimated and the ground truth MIDI files. The alignment accuracy is defined as the percentage of correctly aligned notes whose onset time deviation lies within a certain tolerance window. For ease of use, the implementation in the mir\_eval library \cite{raffel_mireval_2014} is used.
Note offsets are not considered in the evaluation, since they can have greater deviations, caused, for example, by reverb.

\subsection{Baselines}

In order to assess our model's performance, it is compared with the results of using the loosely aligned MIDI and DTW as a state-of-the-art method.
For DTW, the implementation uses the synctoolbox by Müller et al.\ \cite{muller_sync_2021}.
The DLNCO features \cite{ewert_high_2009} are extracted from the audio and the MIDI file respectively and the warping path is then determined by applying the MrMsDTW variant \cite{pratzlich_memory_2016}.
\section{Results}\label{sec:results}

In this section, we evaluate the performance of the proposed method on both the test split of the unaligned MAPS dataset and real-world piano recordings. The alignment quality is assessed using the evaluation metrics defined in Section \ref{sec:metrics}. 
We first compare our model against established baselines, followed by a qualitative example from the MAPS dataset. 
Finally, we conduct an ablation study to analyze the impact of architectural choices, input features, and other key hyperparameters on model performance.

\subsection{Comparison with Baselines}

A comparison of the proposed CRNN model with the baseline methods is shown in Table \ref{tab:comparison}.
For the unaligned MAPS dataset, a maximum timing variation of $\SI{100}{\ms}$ is applied. Therefore, when calculating the metrics on the unaligned MIDI, an alignment accuracy of $\SI{100}{\percent}$ is observed for the $\SI{100}{\ms}$ tolerance window. For more strict windows, the results decrease resulting in a baseline alignment accuracy of $\SI{27.52}{\percent}$ in the $\SI{10}{\ms}$ window.
Using DTW leads to worse performance scores than the unaligned baseline. This is caused by the randomly generated unalignment of the notes, which does not necessarily follow a monotonic warping path.
The proposed CRNN on the other hand leads to an improved alignment and increases the alignment scores significantly. The accuracy for the $\SI{10}{\ms}$ window is increased by over $\SI{10}{\percent}$ and the mean absolute onset error decreases to $\SI{26.33}{\ms}$. 
When training the same CRNN model on a blind transcription task (without MIDI input), performance decreases compared to using MIDI scores.

For the real world piano recordings, the unaligned MIDI files achieve an accuracy of $\SI{20.67}{\percent}$ in the $\SI{10}{\ms}$ tolerance window. By applying DTW this score can be improved to $\SI{41.53}{\percent}$. The proposed CRNN performs slightly worse than DTW for the larger tolerance windows, but performs over $\SI{10}{\percent}$ better in the $\SI{10}{\ms}$ range.

By applying the proposed CRNN on the alignment results of DTW, the overall best results are achieved for the unaligned MAPS dataset and the real world dataset. The results are improved by over $\SI{20}{\percent}$ for the $\SI{10}{\ms}$ tolerance window in comparison to using DTW alone. When comparing the standard deviation for the real world piano recordings, the combined approach also leads to the most consistent results in that case.

\begin{table*}[htb]
	\centering
	\caption{Comparison of our proposed CRNN model with the MrMsDTW using DLNCO features.}
	\label{tab:comparison}
  \begin{tabular}{lccccccc}
    \hline
                                                       & Mean                                            & Median           & Std              & \leqms{100}      & \leqms{50}       & \leqms{25}       & \leqms{10}       \\
    \hline
    \hline
                                                       & \multicolumn{7}{c}{Unaligned MAPS Test Split}                                                                                                                           \\
    \hline
    \hline
    Unaligned                                          & $\SI{28.37}{\milli\second}$                                         & $\SI{24.07}{\milli\second}$          & $\SI{22.51}{\milli\second}$          & $\SI{100.00}{\percent}$         & $\SI{80.74}{\percent}$          & $\SI{57.79}{\percent}$          & $\SI{27.52}{\percent}$          \\

    DTW \cite{muller_sync_2021}                        & $\SI{30.62}{\milli\second}$                                         & $\SI{24.40}{\milli\second}$          & $\SI{25.31}{\milli\second}$          & $\SI{97.74}{\percent}$          & $\SI{79.90}{\percent}$          & $\SI{54.18}{\percent}$          & $\SI{24.79}{\percent}$          \\

    CRNN (Blind Transcription)                               & $\SI{28.07}{\milli\second}$                                         & $\SI{16.84}{\milli\second}$          & $\SI{32.88}{\milli\second}$          & $\SI{95.07}{\percent}$          & $\SI{82.75}{\percent}$          & $\SI{65.86}{\percent}$          & $\SI{36.19}{\percent}$          \\
    CRNN (Ours)                                        & $\SI{26.33}{\milli\second}$                                         & $\SI{15.33}{\milli\second}$          & $\SI{29.03}{\milli\second}$          & $\SI{96.68}{\percent}$          & $\SI{82.69}{\percent}$          & $\SI{66.78}{\percent}$          & $\SI{40.78}{\percent}$          \\

    \textbf{DTW \cite{muller_sync_2021} + CRNN (Ours)} & \textbf{$\SI{25.07}{\milli\second}$}                                & \textbf{$\SI{14.67}{\milli\second}$} & \textbf{$\SI{28.81}{\milli\second}$} & \textbf{$\SI{96.05}{\percent}$} & \textbf{$\SI{84.37}{\percent}$} & \textbf{$\SI{70.12}{\percent}$} & \textbf{$\SI{43.29}{\percent}$} \\

    \hline
    \hline
                                                       & \multicolumn{7}{c}{Real World Piano Recordings}                                                                                                                   \\
    \hline
    \hline
    Unaligned                                          & $\SI{30.56}{\milli\second}$                                         & $\SI{27.78}{\milli\second}$          & $\SI{19.69}{\milli\second}$          & $\SI{98.34}{\percent}$          & $\SI{82.68}{\percent}$          & $\SI{47.56}{\percent}$          & $\SI{20.67}{\percent}$          \\

    DTW \cite{muller_sync_2021}                        & $\SI{16.78}{\milli\second}$                                         & $\SI{12.90}{\milli\second}$          & $\SI{14.86}{\milli\second}$          & $\SI{99.52}{\percent}$          & $\SI{95.09}{\percent}$          & $\SI{78.73}{\percent}$          & $\SI{41.53}{\percent}$          \\

    CRNN (Blind Transcription)                               & $\SI{25.26}{\milli\second}$                                         & $\SI{17.60}{\milli\second}$          & $\SI{26.72}{\milli\second}$          & $\SI{96.78}{\percent}$          & $\SI{87.34}{\percent}$          & $\SI{63.02}{\percent}$          & $\SI{33.30}{\percent}$          \\
    CRNN (Ours)                                        & $\SI{19.03}{\milli\second}$                                         & $\SI{13.44}{\milli\second}$          & $\SI{19.29}{\milli\second}$          & $\SI{98.33}{\percent}$          & $\SI{89.08}{\percent}$          & $\SI{74.74}{\percent}$          & $\SI{53.20}{\percent}$          \\

    \textbf{DTW \cite{muller_sync_2021} + CRNN (Ours)} & \textbf{$\SI{12.33}{\milli\second}$}                                & \textbf{$\SI{7.73}{\milli\second}$}  & \textbf{$\SI{13.56}{\milli\second}$} & \textbf{$\SI{99.54}{\percent}$} & \textbf{$\SI{95.63}{\percent}$} & \textbf{$\SI{86.17}{\percent}$} & \textbf{$\SI{62.02}{\percent}$} \\
    \hline
  \end{tabular}
\end{table*}

\begin{figure*}[htbp]
 \centering
 \providecommand{\fwidth}{.85\textwidth}
 \input{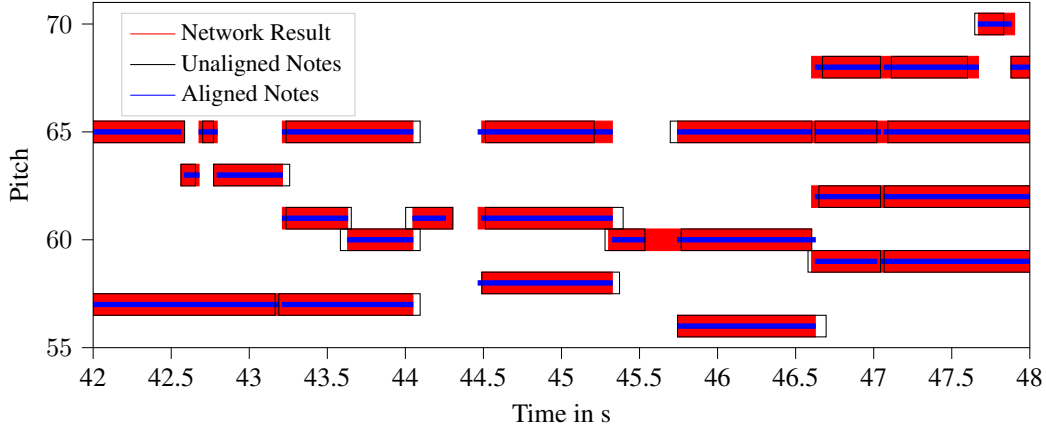}
 \caption{Alignment output of the proposed CRNN model on a MAPS test excerpt. Red blocks indicate the predicted aligned piano roll, blue lines mark ground-truth note onsets, and black boxes denote the original unaligned MIDI. The model effectively adjusts the MIDI timing to match the audio performance, correcting onset misalignments.}
 \label{fig:crnn_example}
\end{figure*}

For a more detailed evaluation, Figure \ref{fig:crnn_example} shows a small excerpt of a synchronization result of the CRNN architecture for an exemplary MAPS test piece. It can be seen that the red coloring, corresponding to the calculated result, often fits the blue line, corresponding to the true value, and corrects the input values represented by the black box. But there are also mistakes visible in this example. It is noticeable that repeatedly played notes are sometimes interpreted as a single, long note. Thus, fewer onsets can be detected, leading to an increased number of false negatives and a lower recall of the network's raw output.
Therefore, the postprocessing step in Section \ref{sec:postprocessing} is used to overcome this issue.

\subsection{Ablation Study}

The following ablation study focuses on the MAPS dataset due to its larger size and greater variability compared to the real-world piano recordings, enabling more meaningful and reproducible evaluations.

\subsubsection{Model Architecture Variations}

Results of different model architecture variations are displayed in Table \ref{tab:architecture_pc}. For the CNN variation, the RNN layer has been removed from the original CRNN architecture. This reduces the model's capabilities of processing the temporal relationship and leads to a significant drop of the alignment accuracies. A direct comparison between Gated Recurrent Unit (GRU) and LSTM cells in the bidirectional RNN layer shows that better results can be achieved when using the LSTM cells.

\begin{table}[H]
	\centering
	\caption{Alignment results on MAPS validation split for various model architectures in percent.}
	\label{tab:architecture_pc}
  \begin{tabular}{lccc}
    \hline
    Architecture & \leqms{50}       & \leqms{25}       & \leqms{10}       \\
    \hline
    CNN          & $77.54$          & $47.17$          & $22.06$          \\
    CRNN (GRU)   & $87.35$          & $73.25$          & $45.94$          \\
    CRNN (LSTM)  & $\mathbf{88.84}$ & $\mathbf{75.25}$ & $\mathbf{47.41}$ \\
    \hline
  \end{tabular}
\end{table}

\subsubsection{Audio Features}
Table \ref{tab:features_pc} shows the impact of different input feature spectrograms on the alignment performance.
Although the calculation of the CQT is not as computationally efficient as the computation of the Mel spectrogram, the CQT comes with the advantage of the same frequency bin arrangement as in the piano roll. Using the logarithmic magnitude scaled CQT leads to an improvement of about $\SI{2}{\percent}$ for the $\SI{10}{\milli\second}$ tolerance window. The improvement for the other tolerance windows is less significant.

\begin{table}[htbp]
  \centering
  \caption{Alignment results on MAPS validation split for various input representations with different magnitude scales in percent.}
  \label{tab:features_pc}
  \begin{tabular}{lcccc}
    \hline
    Input & Scale & \leqms{50}       & \leqms{25}       & \leqms{10}       \\
    \hline
    Mel   & Lin   & $\SI{88.77}{\percent}$            & $\SI{74.07}{\percent}$            & $\SI{45.54}{\percent}$            \\
    Mel   & Log   & $\SI{88.27}{\percent}$            & $\SI{74.00}{\percent}$            & $\SI{44.38}{\percent}$            \\
    CQT   & Lin   & $\SI{88.30}{\percent}$            & $\SI{74.28}{\percent}$            & $\SI{46.20}{\percent}$            \\
    CQT   & Log   & \textbf{$\SI{88.84}{\percent}$}   & \textbf{$\SI{75.25}{\percent}$}   & \textbf{$\SI{47.41}{\percent}$}   \\
    \hline
  \end{tabular}
\end{table}

\subsubsection{Temporal Resolution}
The temporal resolution of the piano roll and the input spectrogram is directly linked to the resolution of the output piano roll. Therefore, decreasing the temporal resolution limits the model's precision when it comes to aligning the MIDI. In Table \ref{tab:resolution_pc}, the impact of varying the number of frames per second (FPS) on the absolute onset error is shown. Increasing the resolution up to $100$ FPS ($\SI{10}{\milli\second}$ per frame) has a positive effect on the performance. Increasing the resolution further does not lead to better results.

\begin{table}[htbp]
  \centering
  \caption{Alignment results on MAPS validation split for different temporal resolutions.}
  \label{tab:resolution_pc}
  \begin{tabular}{lccc}
    \hline
    FPS   & \leqms{50}       & \leqms{25}       & \leqms{10}       \\
    \hline
    $200$ & $\SI{87.65}{\percent}$          & $\SI{74.21}{\percent}$          & $\SI{47.04}{\percent}$          \\
    $100$ & \textbf{$\SI{88.84}{\percent}$} & \textbf{$\SI{75.25}{\percent}$} & \textbf{$\SI{47.41}{\percent}$} \\
    $50$  & $\SI{86.66}{\percent}$          & $\SI{72.11}{\percent}$          & $\SI{42.68}{\percent}$          \\
    $25$  & $\SI{84.20}{\percent}$          & $\SI{61.93}{\percent}$          & $\SI{31.50}{\percent}$          \\
    \hline
  \end{tabular}
\end{table}

\subsubsection{Maximum Human Timing Variations}

For the experiments analyzed up to now, a maximum human timing variation of $\SI{100}{\ms}$ is used in the unalignment procedure of the dataset creation. Experiments with real world musical pieces also yield an average human timing variation of $\SI{95}{\ms}$.
While the extent of human timing variation can depend on the musician's proficiency, studies have shown that such variations are present in both professional and 5-year-old children's performances \cite{penel_sources_1998}. Therefore, different values for the human timing variation have been analyzed in Table \ref{tab:human_error_ae}. There is a correlation between the maximum human timing variation and the alignment performance. Smaller human timing variation results in a lower mean absolute error and a more consistent alignment result.

\begin{table}[htbp]
  \centering
  \caption{Alignment results on MAPS validation split with different maximum human timing variation values.}
  \label{tab:human_error_ae}
  \begin{tabular}{lccc}
    \hline
    Timing Variation           & Mean    & Median  & Std     \\
    \hline
    $\SI{50}{\milli\second}$  & $\SI{18.24}{\milli\second}$ & $\SI{11.38}{\milli\second}$ & $\SI{20.78}{\milli\second}$ \\
    $\SI{100}{\milli\second}$ & $\SI{21.60}{\milli\second}$ & $\SI{12.75}{\milli\second}$ & $\SI{26.16}{\milli\second}$ \\
    $\SI{150}{\milli\second}$ & $\SI{24.15}{\milli\second}$ & $\SI{13.09}{\milli\second}$ & $\SI{30.49}{\milli\second}$ \\
    $\SI{200}{\milli\second}$ & $\SI{25.84}{\milli\second}$ & $\SI{14.19}{\milli\second}$ & $\SI{32.86}{\milli\second}$ \\
    \hline
  \end{tabular}
\end{table}

\subsubsection{Tempo Changes}\label{sec:tempo}

In this experiment, the tempo is changed by scaling the unaligned MIDI with a randomly sampled tempo factor in the range $\left[1 - \text{Max Tempo Factor}, 1 + \text{Max Tempo Factor}\right]$.
With the introduction of tempo changes, the network's capabilities reach their limit. This is displayed in the test results shown in Table \ref{tab:tempo_changes_ae}. It shows decreasing alignment accuracies for increasing ranges of tempo changes.

\begin{table}[H]
  \centering
  \caption{Results on MAPS validation split with tempo changes.}
  \label{tab:tempo_changes_ae}
  \begin{tabular}{lccc}
    \hline
    Max Tempo Factor & Mean    & Median  & Std     \\
    \hline
    $0.00$           & $\SI{21.36}{\milli\second}$ & $\SI{12.40}{\milli\second}$ & $\SI{26.35}{\milli\second}$ \\
    $0.01$           & $\SI{32.77}{\milli\second}$ & $\SI{18.70}{\milli\second}$ & $\SI{36.39}{\milli\second}$ \\
    $0.05$           & $\SI{44.20}{\milli\second}$ & $\SI{24.88}{\milli\second}$ & $\SI{47.80}{\milli\second}$ \\
    $0.10$           & $\SI{47.21}{\milli\second}$ & $\SI{27.54}{\milli\second}$ & $\SI{50.05}{\milli\second}$ \\
    $0.20$           & $\SI{48.61}{\milli\second}$ & $\SI{28.47}{\milli\second}$ & $\SI{50.67}{\milli\second}$ \\
    $0.30$           & $\SI{49.76}{\milli\second}$ & $\SI{30.72}{\milli\second}$ & $\SI{50.59}{\milli\second}$ \\
    \hline
  \end{tabular}
\end{table}

\section{Conclusion}\label{sec:summary}
A novel CRNN based approach for aligning piano rolls with performance audio spectrograms has been proposed. 
In order to synchronize piano recordings with their loosely aligned MIDI annotations, the model has been trained on a dataset compiled by augmenting existing piano transcription examples. To evaluate the model performance on real world data, six real world piano performances have been recorded.
The synchronization system provides an improved alignment accuracy when compared with the unaligned MIDI files. Especially for the very strict tolerance window of $\SI{10}{\ms}$, the sole model outperforms the state-of-the-art approach DTW.
When combining DTW with the proposed CRNN model, the best results can be achieved for the unaligned MAPS dataset and the real world piano pieces.
In an ablation study, several model parameters have been researched and the limits of the model have been tested. The model performs best for the task of finely aligning MIDI notes that have already been loosely aligned in a preceding processing step. Hence, the best results can be achieved when performing DTW in advance.

In future work, this approach could be extended to other instruments using datasets like e.g.\ GuitarSet \cite{xi_guitarset_2018}. Additionally, applying the model during AMT training may improve onset precision. Exploring alternative architectures such as transformers could also further enhance performance.

\bibliographystyle{IEEEtran}
\bibliography{main}

\end{document}